\documentclass[
 aip,
 jap,
 amsmath,amssymb, reprint
]{revtex4-2}

\usepackage{graphicx}
\usepackage{dcolumn}
\usepackage{bm}
\usepackage{hyperref}

\hypersetup{
    colorlinks=true,
    linkcolor=blue,
    filecolor=magenta,
    urlcolor=black,
    citecolor=blue,
   pdftitle={manuscript}
}

\begin{document}

\title[]{Electric field induced negative capacitance in semiconducting polymer}

\affiliation{Department of Physics, Indian Institute of Science Bangalore 560012, India}
\author{Sougata Mandal}
\email{sougatam@iisc.ac.in}
\author{Reghu Menon}

\begin{abstract}

Electric field dependent capacitance and dielectric loss in poly(3-hexylthiophene) are measured by precision capacitance bridge. Carrier mobility and density are estimated from fits to current-voltage and capacitance data. The capacitance varies largely at lower frequency, and it decreases at higher electric fields. The negative capacitance at low frequency and high field is due to the negative phase angle between dipole field and ac signal. The intrinsic carrier density is calculated from fits to the Mott-Schottky equation, and this is consistent with $I-V$ data analysis.  At higher frequency, the carriers do not follow the ac signal and its density drops; and the flat band potential increases mainly due to the build-in potentials within ordered and amorphous regions in the sample.

\end{abstract}

\maketitle


\section{Introduction}

Capacitance as a function of frequency is one of the important dielectric properties of materials related to polarizability, charging, electrical response, etc. Also, it probes the charge storage and discharge capabilities, roles of domains and boundaries, interfacial and bulk polarization \cite{santos2013electric,uludaug2022dielectric,mills2002effects}. In semiconductors, the capacitance measurements are usually used to find out the role of electrode interface, trapping and relaxation processes of charge carriers \cite{doi:10.1063/1.3152797}. The physical and electronic properties of semiconducting polymers very much depend on the chemical processing conditions of the samples. The complex morphology of polymers depends on various processing factors, and this affects the electronic and optical properties, since it is known that trap-states, built-in potential, homogeneity, domain sizes, etc. varies\cite{polym9060212}. In semiconducting polymers, capacitance varies due to several factors like morphology, carrier density, traps, and electrode interface; and its variations can be probed in detail by illumination of light, doping and application of electric field \cite{vollbrecht2022determination,von2019impedance}. Regioregular poly(3-hexylthiophene) [P3HT], with a band gap around 2.2 $eV$, is one of the widely used semiconducting polymers in transistors, solar cells and photodetectors \cite{burgi2002noncontact,chan2010high,burgi2003close,liu2009hybrid,foertig2012shockley,wang2015improved}. The capacitance measurements as a function of frequency and electric field give more insight into the carrier response unlike dc studies \cite{li2011simultaneous}. It is important to study and understand the effect of electric field, especially when semiconducting polymers are used for practical applications like photovoltaic, supercapacitors, batteries, and other electronic devices. Although dc and ac charge transport studies have been reported in P3HT \cite{doi:10.1021/acs.macromol.5b02727,PhysRevB.80.195211,mandal2020impedance,oklobia2018impedance,grecu2004characterization}, precise capacitance measurements in well-ordered samples are lacking. Four-electrode capacitance measurement under d.c. bias is a good technique to understand the electronic properties and characterization of semiconducting polymers for varying frequencies.

In this work, we have shown that the four-electrode capacitance measurement is a very sensitive measurement technique, it can easily probe the processing and morphology dependent roles in the electronic properties of polymer. Especially, parameters like carrier density, built-in potential and interfacial polarization effects that varies with frequency and field are essential fundamental characterization of the semiconducting polymer. A high precision capacitance bridge is used to measure the capacitance and dielectric loss of P3HT films at varying frequencies (50 $Hz$ to 20 $kHz$). This work mainly focuses on how the electric field modifies the carrier density and mobility; and alters the flat band potential and its effect on capacitance, providing insight into the sample's domains, traps and interfaces.

\section{Experimental and Characterization}

Solution processed free-standing films of purified P3HT (Rieke metals, Inc.) of thickness around 20 $\mu m$ are used in this study. The free-standing films are prepared by drop-casting methods on glass substrate with chlorobenzene as solvent. The regularity and uniformity of the sample can be ensured by slow evaporation rate of chlorobenzene, as the sample is prepared in very slow drying condition to enhance the morphological order of the film. We observed that this occurs in samples of thickness of $ 10-20$ $\mu m$ range. Keithley Source-Measure Unit (SMU 2400) is used for dc current-voltage ($I-V$) measurements to find out the charge transport mechanism. Andeen-Hagerling capacitance bridge (Model: AH 2700A) is used in parallel mode configuration to precisely measure the capacitance and dielectric loss values. The capacitance bridge makes the measurements in the form of parallel capacitance ($C_P$) and loss [in terms of conductance, $G=1/R_P$, $R_P$ is the resistance parallelly connected with $C_P$]. Electric field dependent measurements of both current and capacitance are carried out in four-electrode geometry on 2 mm $\times$ 2 mm free-standing films with carbon paint contacts with low contact resistance. The contact pad area is 0.78 $cm^2$. Thin copper wires (20 $\mu m$) are attached to the edges of the sample with carbon paints for applying the electric field to the entire bulk of the sample. For measuring current and capacitance, the wires are glued in the middle top-bottom of the sample with carbon paints, so that the electric field response is measured from all the different morphological domains within the bulk of the sample. Since the sample consists of both ordered and amorphous domains with several interfaces, along with low mobile carriers amidst various trap states, this electrode configuration is quite appropriate to explore the role of in-homogeneous morphology in capacitance, especially for free standing samples. The schematic of the measurement technique is shown in inset of Fig. \ref{fig: current_electric_field}.

\begin{figure}
    \centering
    \includegraphics[scale=0.4]{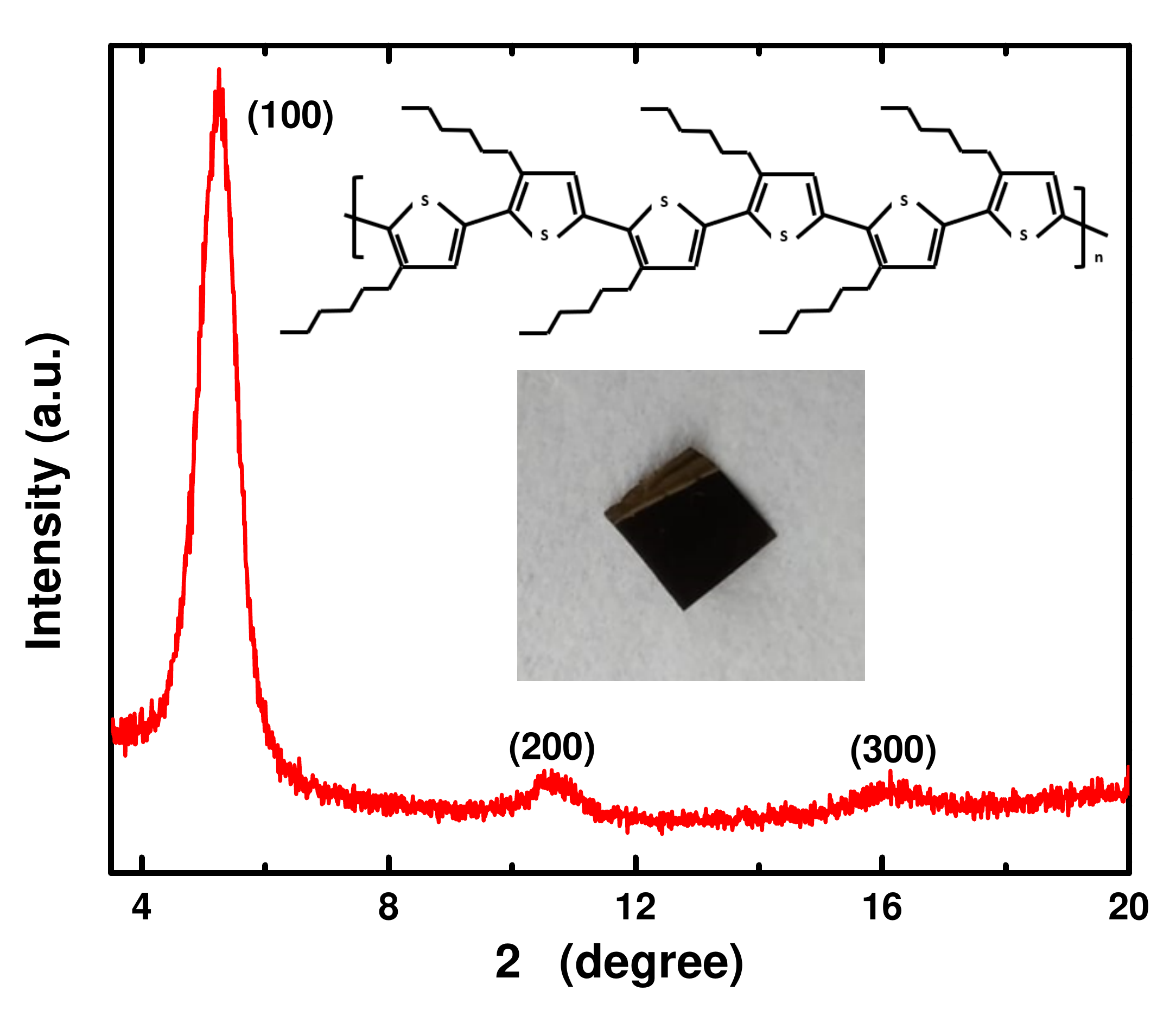}
    \caption{X-ray diffraction (XRD) pattern of regioregular P3HT. Inset shows the P3HT chain and picture of free-standing film}
    \label{fig:xrd_p3ht}
\end{figure}

X-ray diffraction measurement is performed by Rigaku SmartLab high-resolution X-ray diffractometer (XRD) in grazing angle (GI angle = 0.5$^{\circ}$) mode (GIXRD) to access the quality of sample. The grazing-incidence  X-ray diffraction of  P3HT films is shown in Fig. \ref{fig:xrd_p3ht}. The diffraction pattern shows one sharp peak at 2$\theta$ = 5.25$^{\circ}$ and two other peaks at 10.64$^{\circ}$ and 15.96$^{\circ}$ respectively. These peaks are attributed to (100), (200) and (300) planes respectively, which signify the well-organized domains due to stacking of thiophene rings as expected in regioregular structure of polymer chains with ordered hexyl-groups \cite{2006NatMa...5..197K,doi:10.1002/adfm.200400521,winokur1989structural}; hence the quality and morphological order of P3HT film is very good.

\section{Results and Discussion}

\subsection{I-V analysis}

\begin{figure}
    \centering
    \includegraphics[scale=0.4]{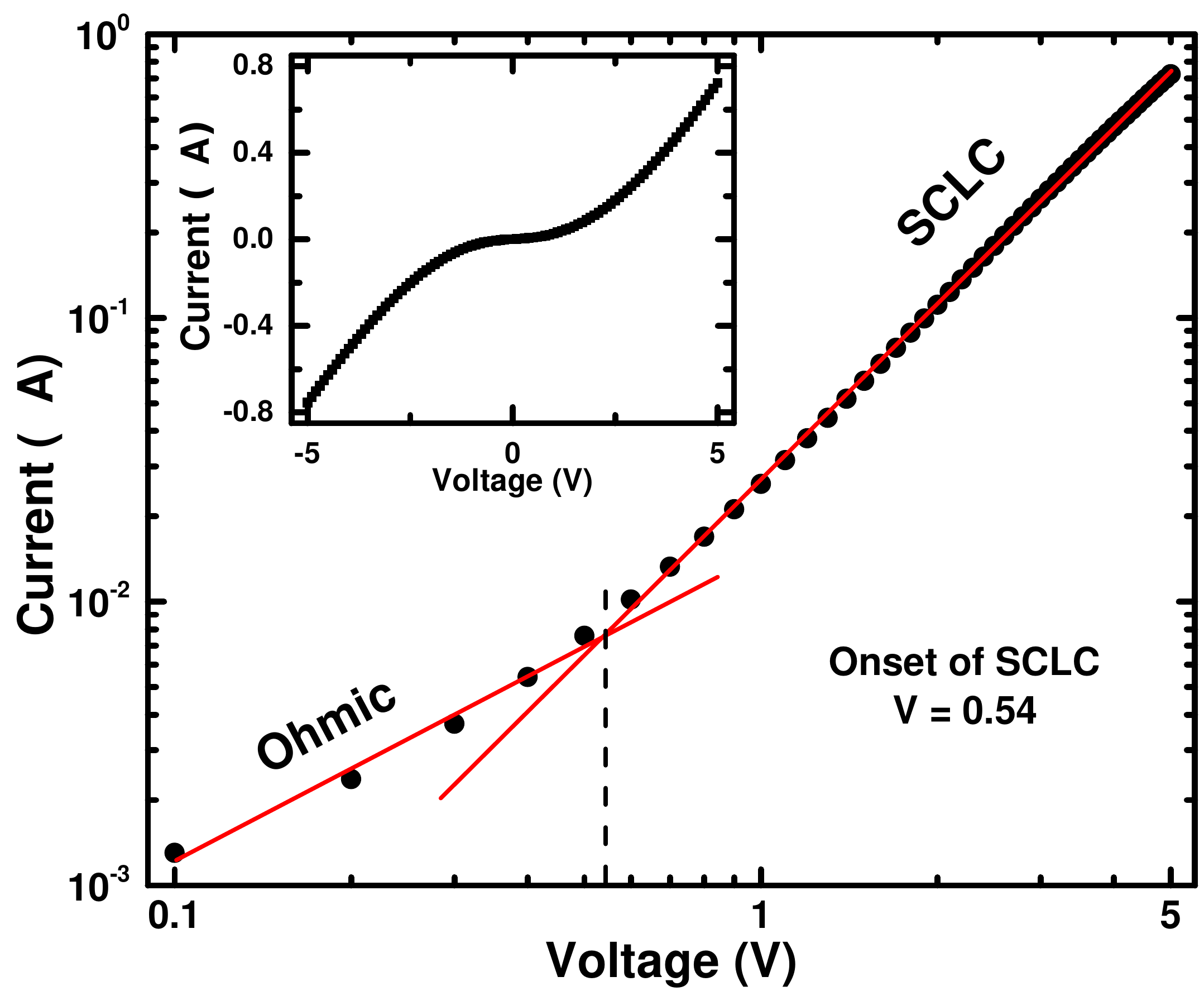}
    \caption{(Color Online) $I-V$ characteristics using two-probe configuration of regioregular P3HT sample in logarithmic scale and data fit to the Ohm’s and the Mott-Gurney SCLC law for lower and higher voltage region respectively. Inset shows the $I-V$ data and onset of SCLC at $V$ = 0.54.}
    \label{fig:IV_two}
\end{figure}

In this high quality P3HT sample, as the structural order is very high (confirmed by XRD data) the delocalized electronic states are quite large. This give rise to better semiconducting properties to the sample, even in thicker films. In our studies, the low field $I-V$ characteristics of free-standing P3HT film is linear, and the measured conductivity of the sample is 5.5$ \times  10^{-6} S/cm$, which is quite good for a semiconductor. Logarithmic plot of $I-V$ characteristics of P3HT film is shown in Fig. \ref{fig:IV_two}, inset shows the $I-V$ data. The lower voltage $I-V$ of the contacts are linear, indicating contact barriers are low, hence the role of contact resistance is negligible. Higher voltage $I-V$ data shown non-linearity due to formation of space-charges within ordered domains, domain boundaries, and ordered-disordered interfaces region. The experimental data are fitted using Ohm’s and Mott-Gurney space charge limited conduction (Mott-Gurney SCLC) laws for two different regions, as shown in the Fig. \ref{fig:IV_two}. The low voltage region (0 $V$ to 0.54 $V$) is mainly dominant by the thermally generated free carriers, and it follows the Ohm’s law [$I = eAN\mu_pV/d$, where $e$: electron charge, $A$: contacts area, $N$: free carrier density, $\mu_p$: mobility of carriers, $V$: applied voltage and $d$: sample thickness]. For higher voltage region (0.54 $V$ to 5 $V$), injected charge carriers are forming space-charge regions and the transport is followed by the Mott-Gurney SCLC [$I = 9\epsilon\mu_pV^2/8d^3$, where $\epsilon$: dielectric permittivity] \cite{mott1948electronic,gurney1964electronic,kao1981electrical}. The charge carrier mobility and carriers concentrations are calculated by equating two equations at $V$ = 0.54 , and this gives $\mu_p = 7.40 \times 10^{-2} cm^2/V.s$ and $N = 3.15\times 10^{11} /cm^3$, respectively \cite{lampert1970current,doi:10.1063/5.0042737}. The carrier density, space charge formation and its variations with electric field as a function of frequency is further discussed in the electric field dependent analysis section.

\subsection{Electric field dependent analysis}

\begin{figure}
    \centering
    \includegraphics[scale=0.4]{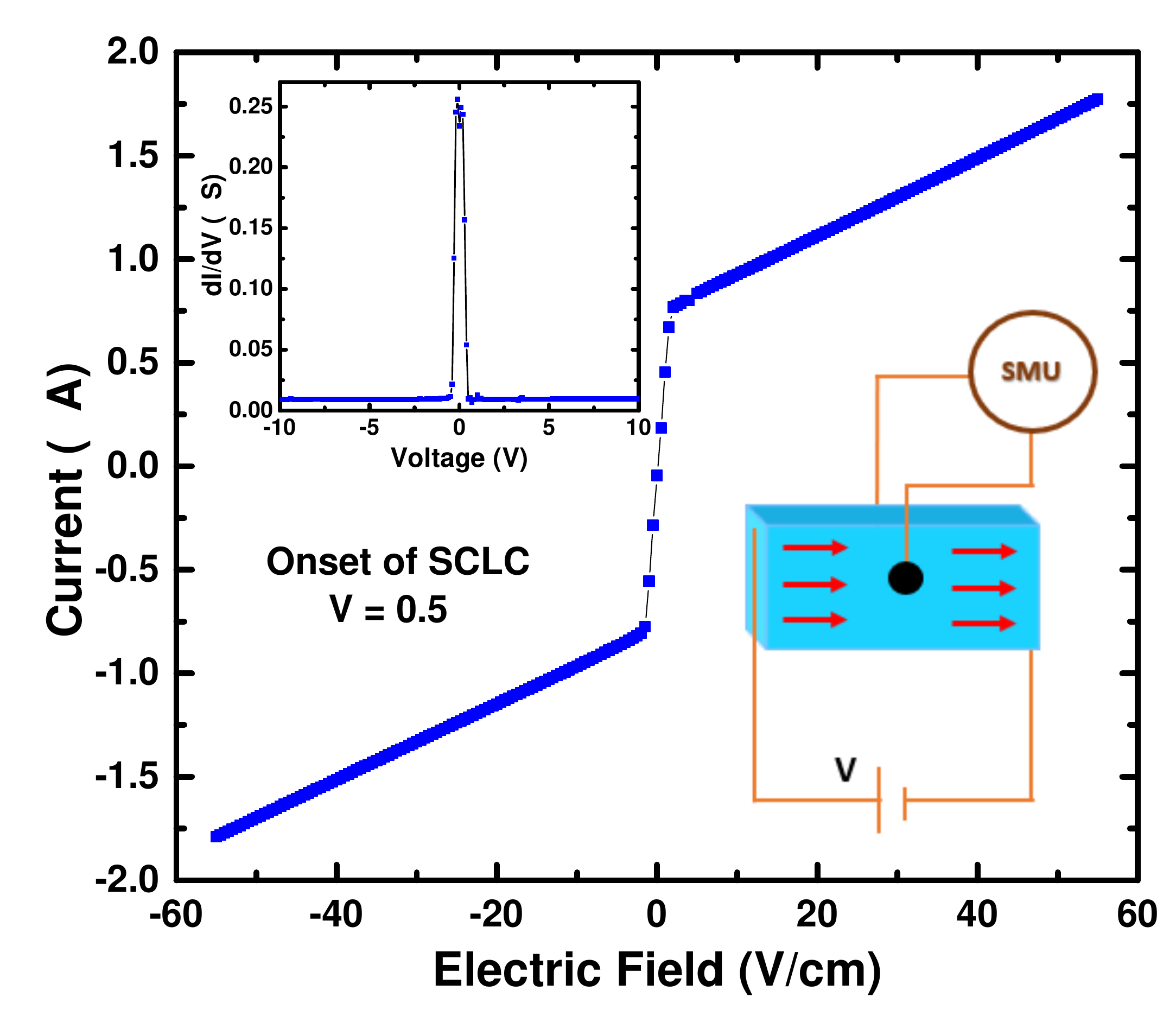}
    \caption{(Color Online) The variation of current with electric field of regioregular P3HT sample. Inset shows schematic of measurements and the differentiation of current for varying voltage ($dI/dV$). It shows onset of SCLC is at $V$ = 0.5}
    \label{fig: current_electric_field}
\end{figure}

The variation of current, in this four-electrode configuration, with applied electric field is shown in Fig. \ref{fig: current_electric_field}. The initial sharp increase in low field is due to thermally generated carriers. The slope change at higher field is due to the onset of space charge formation as the rate of injected charge carrier increases. The combination of low mobility and traps in interface regions facilitates the formation of space charge. The inset in Fig. \ref{fig: current_electric_field} is to highlight the onset of SCLC from Ohmic conduction at $V$ = 0.5. This is being quantified by how much the value of $dI/dV$ has changed. The value of $dI/dV$ increases by a factor of 22, which is quite large, shows the dominance of SCLC. This value is quite consistent with the observation of the onset of SCLC of the two probe $I-V$ measurements ($V$ = 0.54) as in Fig. \ref{fig:IV_two}. This increase in current with applied field shows that the four-electrode configuration is quite appropriate to find the correlation among the morphology and charge transport in semiconducting polymers. Hence, similar four-electrode configuration is used in the field dependent capacitance measurements. Since electric field is applied all through the bulk of the sample, the capacitance measurements across the thickness can give insight into the carrier density, flat band potential, etc. As the field dependent capacitance measurement is more sensitive, the role of in-homogeneous morphology in charge transport can be explored in more detail as discussed below.

Same sample with the above electrode configuration is used in the field dependent capacitance measurements. The applied dc voltage creates electric field within the bulk of the sample, and the capacitance is measured by applying  small ac field in the perpendicular direction. The constant dc electric field creates polarization within the sample, and this is varied as we increase the applied electric field; and the variation is being probed by capacitance-frequency measurements. Moreover, the amplitude of ac field is kept very low to work in linear regime, and the applied dc field is varying from low to high-fields.

\begin{figure*}
    \centering
    \includegraphics[scale=0.6]{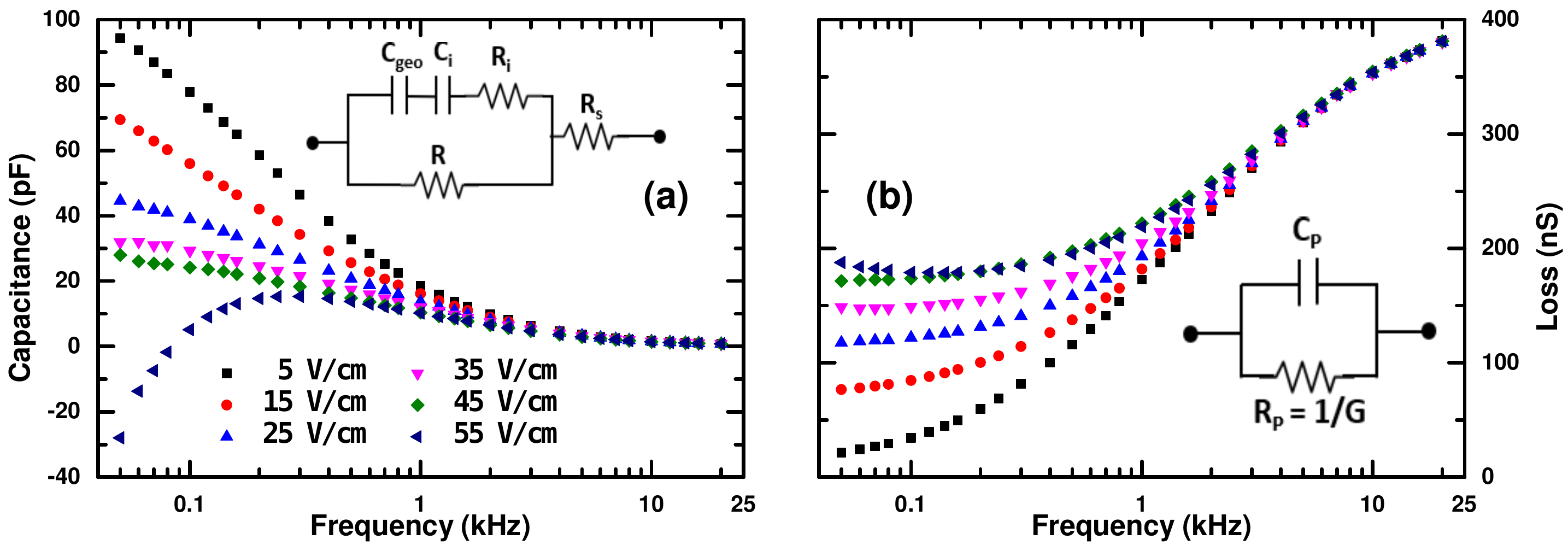}
    \caption{(Color Online)  Electric field dependent (a) capacitance (b) dielectric loss values of regioregular P3HT at varying frequencies. The same schematic is used for capacitance measurement by varying electric field as shown in inset of Fig. \ref{fig: current_electric_field}; Inset shows the equivalent model [in Fig. (a)] and instrument measurement mode [in Fig. (b)]}
    \label{fig:capacitance_loss}
\end{figure*}

Capacitance in this semiconducting polymer is mainly arises due to the accumulation of charges in domain boundaries, traps, and sample-electrode interfaces. In P3HT polymer matrix, the co-existence of amorphous and crystalline regions (see Fig. \ref{fig:xrd_p3ht}) creates interfacial polarizations, hence the measured capacitance is different from geometrical calculated values. Geometrical capacitance [ $C_{geo}$] is defined as, $C_{geo}=\epsilon A/d$ ($\epsilon = k\epsilon_0, k$: dielectric constant).  Dielectric loss is mainly due to the polarizations within the sample. The loss of energy is caused by the movement of charges, as the alternating electromagnetic field switches the direction of motion of carriers. In P3HT, the capacitance and dielectric loss depend on many factors like crystallinity, traps distribution, sample-electrode interface trap states, carrier density and mobility.

The variation of capacitance and dielectric loss with frequency at different electric fields are shown in Fig. \ref{fig:capacitance_loss}(a, b). Electric field varies from 5 to 55 $V/cm$ in steps of 10 $V/cm$. The capacitance decreases significantly at lower frequency (50 $Hz$ to 2 $kHz$), while at higher frequencies (2 $kHz$ to 20 $kHz$) the variation is less. The low frequency decrease in capacitance with electric field can be explained by using space charge effect. The capacitance is mainly due to the geometrical contributions and interfacial polarizations, in which the former is not field-dependent. The calculated $C_{geo}$ is 122.43 $pF$ (considering  $k$ = 3.75 \cite{wang2018high}, $\epsilon = 3.75\epsilon_0$); this value is constant as geometric capacitance dependent parameters like film thickness, contact area and dielectric constant of the film are unchanging during measurements. $C_i$ is the capacitance mainly due to the charges at domain boundaries and sample-electrode interfaces, and these varying as a function of frequency and applied electric field. The interfacial polarizations which are mainly due to the co-existence of ordered and amorphous regions and the interfaces within the sample, and this is the dominant part in the measured capacitance, and this varies with applied field. The sample electrode interfacial polarization can also contribute, especially at high field. These are the major contributions to the capacitance. $R_P$ is the summative resistive contribution mainly due to the polarizations (interfacial: $R_i$, bulk: $R$) of the sample and contact resistances ($R_S$). The effective capacitance [$C_P=C_{geo} C_i/(C_{geo}+C_i$)] and dielectric loss in the form of conductance ($G=1/R_P$) are measured with capacitance bridge, as shown in the insets of Fig. \ref{fig:capacitance_loss}(a, b). From Fig. \ref{fig:capacitance_loss}(a), for low applied electric field, injected charge carriers diffuse rapidly and reduce the local charge carrier density variations within domains and domain boundaries.Hence the polarization effects due to injected carriers are negligible, as the capacitance consistently decreases with the applied field. As the value of electric field increases this reduction of capacitance value increases even more, as shown in Fig. \ref{fig:capacitance_loss}(a). At 45 $V/cm$, the variation of capacitance is minimal with frequency, suggests that the local variations of the dielectric properties from injected carriers are getting saturated at higher fields. As we scaled the dielectric loss in the form of conductance ($G = 1/ R_P$), the value of conductance is increasing. The data show that the dielectric loss (resistance part) increases (decreases) with electric field, and this shows that the injected carriers are diffusing and contributing to the enhancement in conductance, [see Fig. \ref{fig:capacitance_loss}(b)]. A negative capacitance value is observed at low frequencies upon further increase in the electric field (55 $V/cm$). At high electric fields, large number of charge carriers cannot homogeneously diffuse throughout the sample; and this results in the formation of space-charges in domains, boundaries and electrode-interface regions. At low frequencies, charge carriers have sufficient time to accumulate in domains and interface regions, and this results in local built-in potentials which alters the small ac signal in capacitance measurements. Dipoles due to interfacial polarizations find it difficult to synchronize with the small ac signal at lower frequencies, and this results in a negative phase angle between dipole electric field and the applied ac signal\cite{guan2017space}. This gives rise to the observed negative capacitance. At higher frequencies, the ac signal changes quickly, and the accumulation of charge carriers becomes less dominating factor, and the negative capacitance effect cannot be observed. The measured corresponding dielectric loss (conductance) increases as the electric field increases, as in Fig. \ref{fig:capacitance_loss}(b). When the injected charge carriers increase with electric field, the sample resistance decreases, resulting in higher dielectric loss (conductance). At high applied electric field (55 $V/cm$), the loss value (conductance: 186 $nS$ or resistance: 5.37 $M\Omega$, 50 $Hz$) is present, as there still exists a non-zero polarization and this leads to the dielectric loss. The rather smooth and systematic variation of capacitance with field is consistent with the models, and this further authenticates the measurement method, as discussed below section.

\begin{figure}
    \centering
    \includegraphics[scale=0.35]{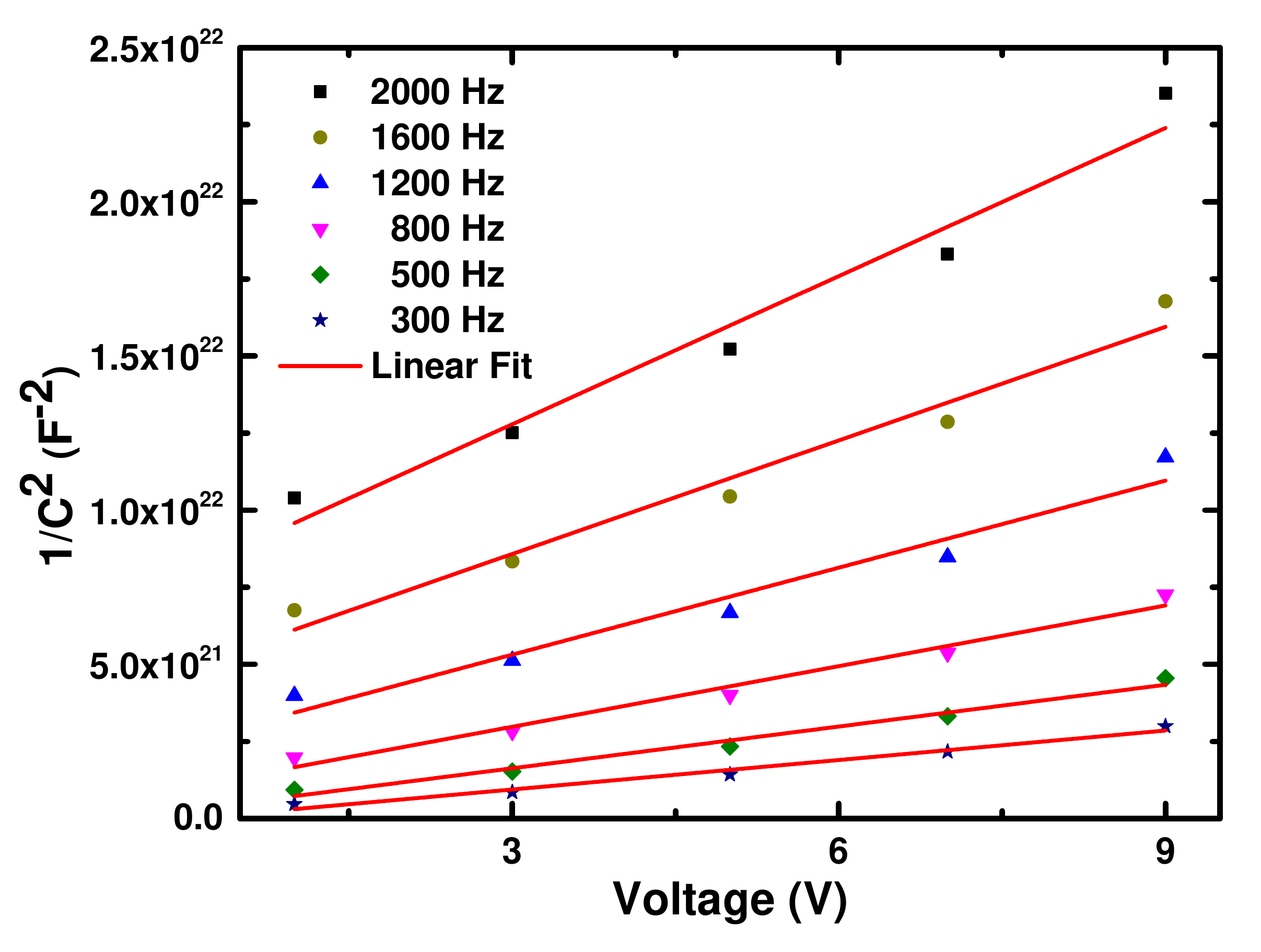}
    \caption{(Color Online) $1/C^2$ vs. Voltage plot for various frequencies and data fit to the Mott-Schottky equation.}
    \label{fig:mott_schottky}
\end{figure}

The measured capacitance in P3HT is characterized by the Mott-Schottky equation \cite{windisch2000mott,doi:10.1021/acs.jpcc.9b09735}, as given by equation \ref{equ:mott_equ},

\begin{equation}
   \frac{1}{C^2} = \frac{2}{e\epsilon A^2 N} (V-V_{FB}- \frac{k_B T}{e})
   \label{equ:mott_equ}
\end{equation}

where, $C$ is the measured capacitance at potential value $V$, $V_{FB}$ is the flat band potential, $k_B$ is the Boltzmann constant, $T$ is the absolute temperature. $1/C^2$ vs. $V$ is plotted for different frequency values, and a linear fit is obtained by using the Mott-Schottky equation as shown in Fig. \ref{fig:mott_schottky}. It is observed that as frequency increases both the slope of the linear fit and intercept value increases. From the slope and intercept value of the fit, free carrier density and flat band potential can be calculated, which are given by equations \ref{equ:mott_equ1} and \ref{equ:mott_equ2}, respectively:

\begin{equation}
   N=\frac{2}{e\epsilon A^2} \times \frac{1}{(Slope)}
   \label{equ:mott_equ1}
\end{equation}

\begin{equation}
   V_{FB} = -[\frac{Intercept}{Slope} + \frac{k_B T}{e}]
   \label{equ:mott_equ2}
\end{equation}

\begin{figure}
    \centering
    \includegraphics[scale=0.3]{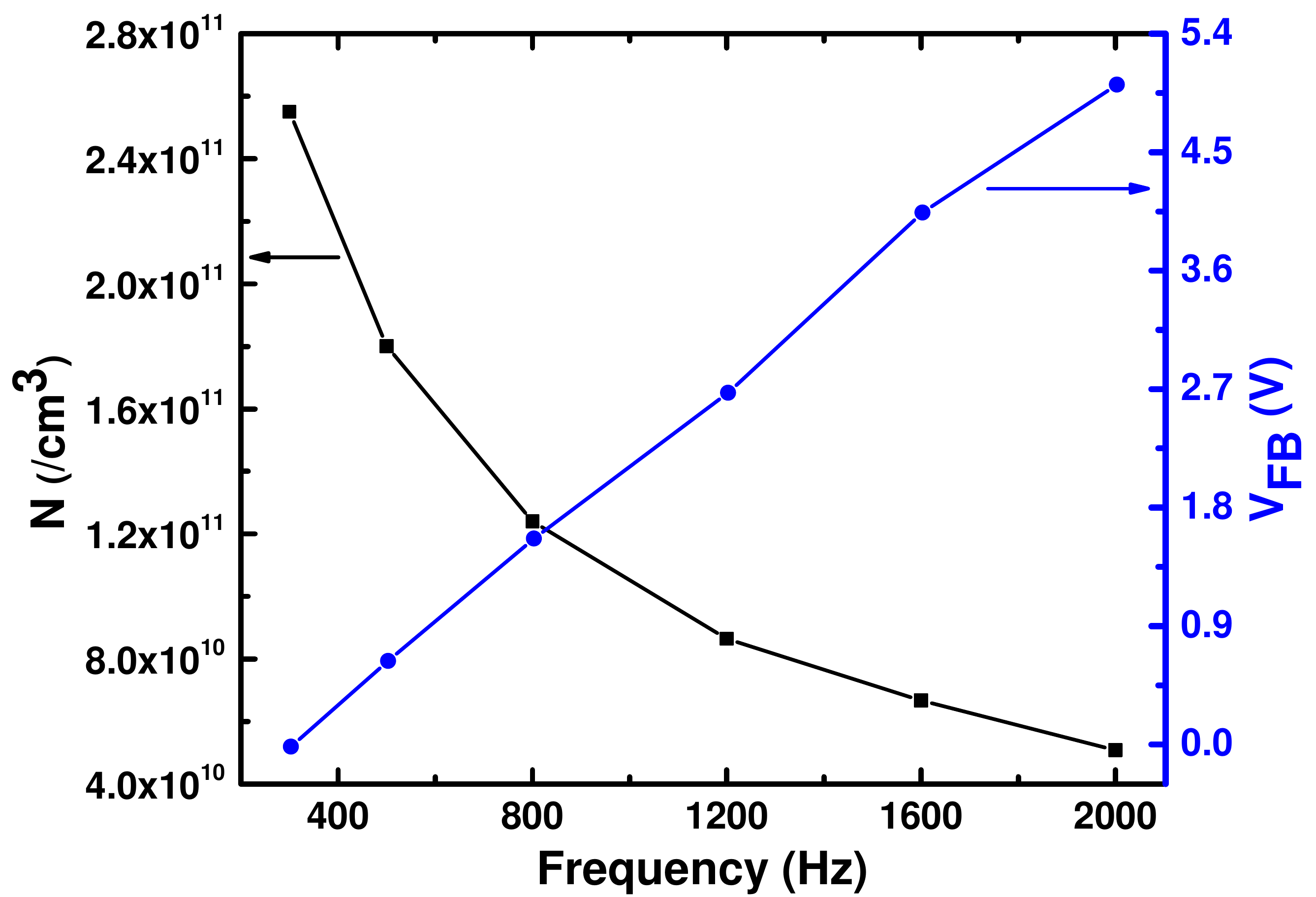}
    \caption{(Color Online) Variation of carrier density (left axis) and flatband potential (right axis) with frequency.}
    \label{fig:carrier_density_flatband}
\end{figure}

The frequency dependent free carrier density and flat band potentials are plotted, as shown in Fig. \ref{fig:carrier_density_flatband}. The calculated free carrier density at 300 $Hz$ is 2.55 $\times 10^{11}/cm^3$. This carrier density is nearly consistent with the value obtained from the fit to $I-V$ data (See Fig. \ref{fig:IV_two}), as discussed in earlier section. The carrier density decreases to 5.08 $\times 10^{10}/cm^3$ at higher frequencies and this is due to the trapping of carriers. De-trapping of the carriers cannot be accomplished at smaller time scales at higher frequencies, and the carriers lag-behind in following the ac signal. This indicates that the relaxation processes take longer time at higher frequencies, especially as the carrier density decreases due to trappings. The flat band potential increases with frequency to 5 $V$ as shown in Fig. \ref{fig:carrier_density_flatband}. The measured system is symmetric and hence the dc flat band potential is zero; in our analysis also the low frequency flat band potential is almost zero (at 300 $Hz, V_{FB} \approx 0)$; it increases to $5 V$ in higher frequency.  At lower frequencies, injected charge carriers are diffusing rapidly throughout the sample as carriers get sufficient time and the resulting potential difference between ordered and amorphous region is less, and this results in lower flat band potentials, as in Fig. \ref{fig:carrier_density_flatband}. As frequency increases, the associated time scale decreases and the charge carriers tend to accumulate at ordered and amorphous interface regions. This creates build-in potential differences within the sample and an increase in flat band potential, as determined from the Mott-Schottky plot analysis. Also, the trapping-detrapping processes involved in the charge transport, as the applied field varies, contributes to this build-in potential; and this in turn depends on the nanoscale morphology of the sample.

Hence, the variation of capacitance as a function of the electric field and built-in potential as a function of frequency can be used to probe the variations in nano-morphology and its role in charge transport in samples prepared under various conditions. This interesting observation in the field induced negative capacitance in semiconducting polymers also gives information about the role of morphology in charge transport. In earlier report, the negative capacitance observed in ferroelectric field effect transistors (FeFET) can be used in non-volatile and flash memory applications\cite{peng2019nanocrystal,kim2022opportunity}. This study has shown that the field induced negative capacitance in semiconducting polymers, especially at lower frequency and higher fields, has potential applications in negative capacitance-based devices.

\section{Conclusions}

In summary, charge carrier mobility and concentration are calculated from current-voltage measurements in semiconducting polymer P3HT. This work has shown precise capacitance measurements as a function of electric field can provide a comprehensive electrical characterization of semiconducting polymer, especially from the variations in capacitance and dielectric loss. The capacitance values reduce as the electric field increases, especially at lower frequency range. The negative capacitance at lower frequency and higher fields gives insight into space charge formation and the role of traps in transport. The Mott-Schottky model analysis shows that as frequency increases the free carrier density decreases and flat band potential increases. This indicates that the charge carriers are weakly following the ac signal at higher frequencies. The Mott-Schottky analysis at varying frequencies can be used for the electrical characterization of semiconducting polymers and devices.

\begin{acknowledgements}

Authors would like to thank Mr. Anindito Bose (Aimil Ltd.) for the Capacitance Bridge (Model AH2700A) instrument. SM acknowledges Department of Science and Technology, India for INSPIRE fellowship.

\end{acknowledgements}

\section*{REFERENCES}

\bibliography{references}.

\end{document}